\documentclass[prl,preprint,showpacs,preprintnumbers,amsmath,amssymb,floatfix]{revtex4-1}
\usepackage{graphicx}
\usepackage{dcolumn}
\usepackage{units}
\usepackage{bm}
\usepackage{amsfonts}
\usepackage{amssymb}

\begin{document}
\title{Plasmon electron-hole resonance in epitaxial graphene}
\vspace{2cm}
\author{C. Tegenkamp$^1$\email{tegenkamp@fkp.uni-hannover.de}, H. Pfn\"ur$^1$, T. Langer$^{1,2}$, J. Baringhaus$^1$,  and H. W. Schumacher$^2$}
\affiliation{
$^{1}$Institut f\"ur Festk\"orperphysik, Leibniz Universit\"at Hannover, Appelstrasse 2, D-30167 Hannover, Germany\\
$^{2}$Physikalisch-Technische Bundesanstalt, Bundesallee 100, D-38116 Braunschweig, Germany}
\date{\today}

\begin{abstract}
The quasiparticle dynamics of the sheet plasmons in
epitaxially grown graphene layers on SiC(0001) have been
studied systematically as a function of temperature,
intrinsic defects, influence of multilayers and carrier density. The opening of the inter-band loss
channel appears as a characteristic  upward shift in the
plasmon dispersion and a dip in the width of the loss peak,
which is explained as a resonance effect in the formation
of electron-hole pairs. Despite the existence of strong
electronic correlations, the plasmon dispersion can  be
quantitatively described within the framework of a nearly
free electron gas.
\end{abstract}
\pacs{81.05.ue, 79.20.Uv, 73.20.Mf} \maketitle

Graphene, as a two-dimensional (2d) lattice of $sp^2$--hybridized carbon atoms,
proves to be  a  model system to study fundamental
questions regarding electronic correlations, collective phenomena,
many-body interactions, dynamical processes and their interrelations
\cite{Novoselov04,Geim07,Bostwick10}. Single  layers can be synthesized by different techniques, e.g.
exfolation of graphite or decomposition of hydrocarbons
on transition metal surfaces
\cite{Michely08,Sutter08}. Well orientated
graphene layers on insulating substrates can be grown
on SiC substrates by sublimation of Si. While the
morphology, the interface, the electronic structure, and
even transport properties have been studied extensively in the past
\cite{Ohta06,Riedl07,Emtsev08,Langer09,deHeer08},  plasmons in graphene have attracted notice only recently
\cite{Liu08, Liu10, Langer10}. As we will show, for studies of this prototype of  an excited collective state in 2d, graphene is indeed a perfect model system.

Although plasmons are used in nano-scaled systems, e.g,  for opto-electrical conversion
and resonant energy transfer processes  \cite{Atwater10},  the prospects of plasmons  of   2d films or even of atomic wires are still  at its beginning  \cite{Diaconescu07,Rugeramigabo10}.  The comparatively flat dispersion in these systems provides ultrashort plasmon wavelengths. Thus, confinement on the nanometer scale becomes feasible, but  strong damping of plasmon excitations and their sensitivity to defects on the atomic scale are challenges to cope with.  Therefore, these studies require geometrically perfect and at the same time electronically flexible materials, such as graphene.

On the other hand, the present knowledge on 1d and 2d sheet plasmons is quite limited and, in particular,  the influence of defects and the interaction with other quasi-particles is barely understood.   It is clear that plasmon damping becomes  dominant once the plasmon dispersion curve intersects the single particle band leading to Landau damping by single particle   intra-band excitations ($SPE_{intra}$) as seen, e.g., in Ag/Si(111) \cite{Nagao01,Nagao07}.
Outside the Landau regime, this mechanism is still effective, but cannot
be described without explicit considerations of correlations \cite{Hwang07,Hwang09,Hill09}
or momentum transfer processes \cite{Langer10}.
Energetically low lying valence band states are considered only  in terms of a
polarizable background  causing a quasi-linear dispersion at long wavelengths \cite{Diaconescu07}.
However, couplings to other particles are neither included in random phase based descriptions \cite{Hwang07},
nor were signatures of resonant coupling seen so far in experiments. Only for hybrid systems,  where plasmons
interact  with  molecular orbitals or excitonic states
\cite{Ni10}, a resonant coupling was recently reported.

In this Letter the sheet plasmon dispersion of graphene and the  coupling
to various decay channels is
analyzed in detail by  angle resolved high resolution
electron energy loss spectroscopy (EELS).
As we will show, once inter-band transitions are allowed by energy and momentum
conservation ($SPE_{inter}$),  resonant coupling of this new decay channel  leads to  pronounced de-tuning of the
plasmon frequency around Fermi energy and to an enhanced damping.
The relevance of a similar type of
quasiparticle dynamics in graphene, namely resonant coupling between plasmon and hole,
called plasmaron,  during photoemission was recently observed  by
Bostwick et.al. \cite{Bostwick07,Bostwick10}.  It corresponds to a complementary process
to that studied here,  and demonstrates the importance of
resonant couplings for the plasmon decay in graphene.
Thus, our findings  broaden the standard
picture for plasmon excitations and damping mechanisms.

%---------- Experimental details----------------

The growth of graphene and the  measurements were performed
under ultra high  vacuum (UHV) condition. As substrate,
Si-terminated 6H-SiC(0001) samples (n-doped, $\approx
10^{18} cm^{-3}$ from SiCrystal AG) were used and graphene
was grown by sublimation of Si while annealing the SiC
substrate to approximately 1500~K. Further details  can be
found in Refs.~\onlinecite{Langer09,Langer10}. The plasmons
were measured by using  a combination of high resolution
electron loss spectrometer (EELS) as electron source and
detector with a low energy electron diffraction system
(LEED) providing simultaneously high energy and momentum ($k_{||}$)
resolution \cite{Claus92}. Typical operating parameters
were 25 meV energy resolution at a $k_{||}$ resolution of
$\rm 1.3 \times 10^{-2}~\AA^{-1}$.
%---------------Results and discussion------------------------
\begin{figure}[tb]
\includegraphics[width=0.9\columnwidth]{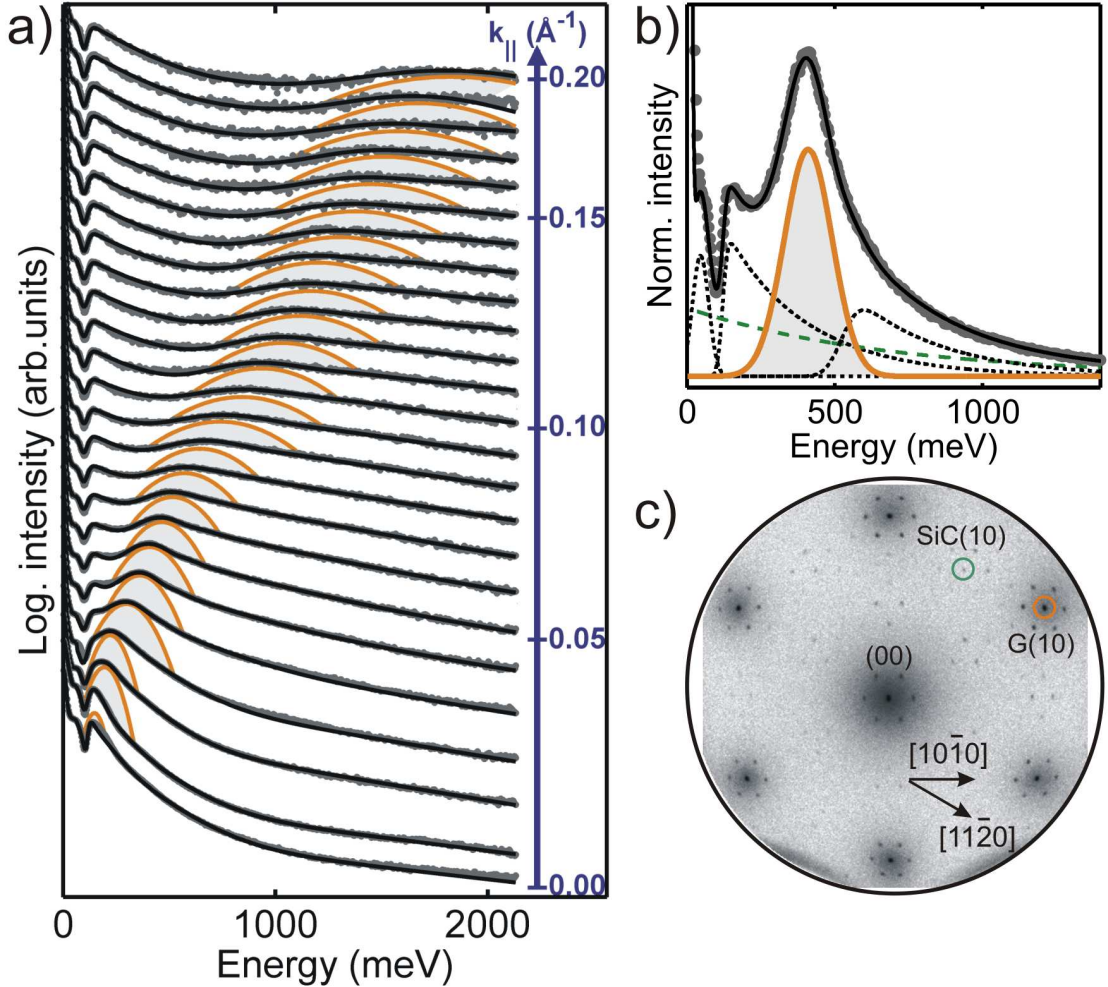}
\caption{\label{FIG1} (color online) a) Loss spectra (semi-log scale) for
1~ML graphene  grown on H--etched SiC(0001) as a function of momentum transfer, $k_\parallel$.
Primary electron energy 20~eV. Only
the contribution  of the 2d sheet plasmon to  the loss
spectrum is highlighted. b) Details of the fitting model, shown exemplarily for $k=0.05~\AA^{-1}$,
containing, apart from the main plasmon peak (solid line (orange)), a Drude tail (dashed (green)),
phonon- and multiplasmon losses (dotted lines). For details of the fitting procedure see Ref.~\cite{Langer10}.
c) LEED pattern of 1~ML of graphene grown on SiC(0001) (E=140eV).}
\end{figure}

A sequence of EEL--spectra of a monolayer (ML) of graphene taken at different $k_{\|}$
values along the $\Gamma$--K direction is shown in
Fig.~\ref{FIG1}. The broad loss peak can unambiguously be
attributed to the sheet plasmon in graphene \cite{Liu08,Langer10}. A detailed analysis of the loss peaks
(see Fig.~\ref{FIG1} and Ref.~\cite{Langer10})
reveals that the loss spectra can be  decomposed into the main plasmon loss,
two low energy phonon losses at 70~meV and 150~meV
without significant dispersion, and small contributions from higher excitations, e.g,  multipole plasmons.
Already from Fig.~\ref{FIG1}a) it is obvious that the plasmon losses reveal
two main characteristics: the halfwidth of the loss,
already large at low $k_\parallel$, suddenly increases
strongly around $\rm 0.1$~\AA$^{-1}$ (cf. with
Fig.~\ref{FIG4}). Secondly, also the peak position shifts
more strongly in this range of momentum.
Both features are directly linked to the opening of a new decay channel by inter-band
e-h excitations, as we will show.

\begin{figure}[tb]
\includegraphics[width=.6\columnwidth]{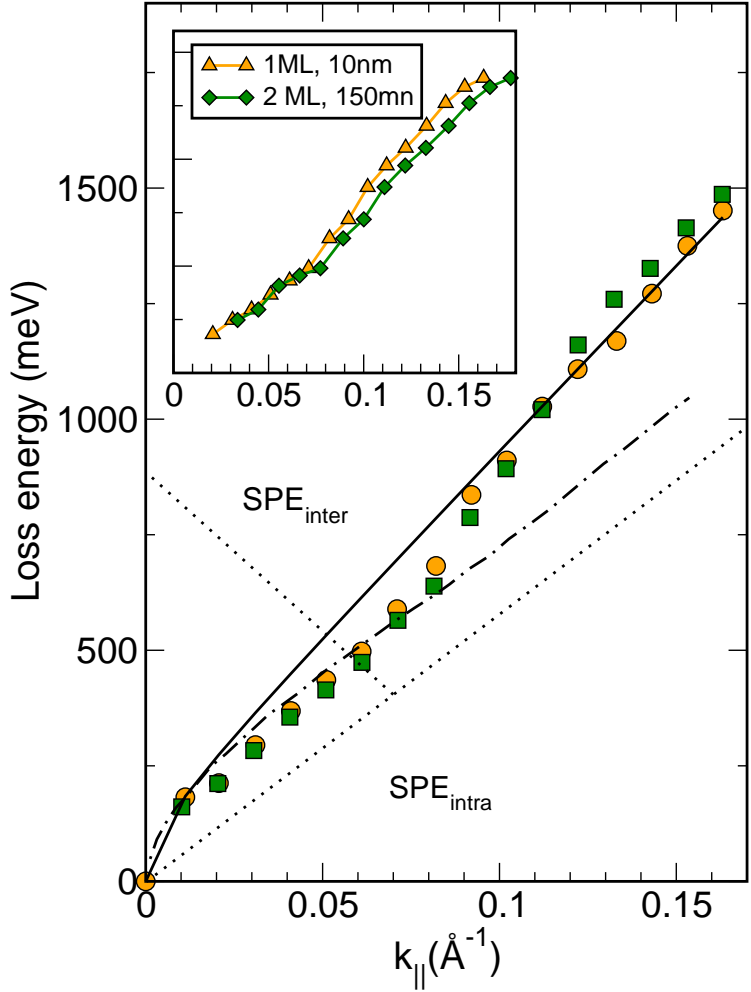}
\caption{\label{FIG2} (color online) Plasmon dispersion for
1~ML graphene layer (average terrace length
150~nm) grown on H--etched samples, measured at 300~K
($\bigcirc$) and 80~K($\Box$). Fitted theoretical curves: nearly
free electron gas ($N=1 \times 10^{13}$cm$^{-2}$, 0.06~m$_e$, \cite{Stern67}) and with
dynamical screening \cite{Hwang07} (dashed-dotted). Dotted lines: Boundaries for
intra- and inter-band SPEs derived from the graphene band
structure. Inset: Independence of dispersion on
step density (examples shown with terrace widths of 10nm and 150 nm)
and on interaction to multilayers.}
\end{figure}

For this purpose, we first discuss the plasmon
dispersion,  shown in Fig.~\ref{FIG2}, revealing a $\sqrt{k_\parallel}$ dependence at small $k_\parallel$.
It rises almost linearly up to $\rm k_F
=0.07 \pm 0.01~\AA^{-1}$,  but then increases strongly again at higher $k_\parallel$ values,
very much in contrast to other known plasmon dispersions in 2d systems
\cite{Nagao01,Rugeramigabo08}. This dip in
the plasmon dispersion is an  intrinsic property of
pristine graphene, and does not
depend on defect concentration, coupling
to  graphene multilayers or on temperature. Therefore, as seen from the coincidence of the dispersions
measured at 80~K  and at 300~K (Fig.~\ref{FIG2}), an increase of the carrier
concentration by thermal activation can be ruled out as well
as the coupling to high-frequency phonons.
Although defects, e.g., atomic steps,  influence sensitively the lifetime of the
plasmons, particularly outside  the Landau damping regime
\cite{Langer10}, the characteristics of
dispersion is  again unaffected,  as exemplarily shown  in the
inset of Fig.~\ref{FIG2}. Even in presence of
multilayers the dip is visible and unshifted.
Only for large k--values a
small redshift due to the high polarizability  of the
graphene layer underneath is found, which is consistent with
Ref.\onlinecite{Liu10}.
The position of the dip in the
dispersion does neither coincide with
step periodicities nor with potential modulations caused
by the superlattice of the bufferlayer underneath
\cite{Kim08}. Hence, so-called zone boundary collective
state effects \cite{Foo68}, as seen, e.g., in Al
films \cite{UrnerWille77}, are not crucial in graphene.
Since the filling factors are comparably low, also
intervalley scattering between Dirac cones can be excluded
\cite{Hill09}. The dip position, however, is very close to the
Fermi wave vector at the given doping concentration ($E_F \approx 400 meV$).

Interestingly, existing theoretical models describe only sections of this dispersion.
E.g., the nearly free electron gas model
(NFEG) with inclusion of first order non--local field
effects, as derived by Stern \cite{Stern67}, describes the experimental
data well in the limit of large momentum transfer ($\rm
k>0.1~\AA^{-1}$) assuming an electron density of
$N=1\times10^{13} 1/cm^2$ and an effective mass of
$m^*=0.06\pm 0.01~m_e$ (solid line in Fig.~\ref{FIG2}).
These values are close
to those deduced from photoemission \cite{Ohta06}.
Inclusion of dynamic screening for both electrons
and holes (dashed--dotted
line, \cite{Hwang07}) shifts the dispersion to lower
values and now yields a perfect match to experiment for small
$k_\parallel$ up to $\rm 0.09~\AA^{-1}$, but underestimates the plasmon
energies in the $SPE_{inter}$ regime. Although  e-h excitations are included in
these RPA descriptions, the NFEG models do not contain effects like resonant
energy transfer with corresponding k-dependent changes of lifetimes.
As we will show below, only the inclusion of resonant energy
transfer at $E_F$ due to opening of the inter-band e-h excitations channel
allows a quantitative fit of the measured
dispersion curve, now even within an extended NFEG model.
%------ Damping model  ---------

\begin{figure}[tb]
\includegraphics[width=0.9\columnwidth]{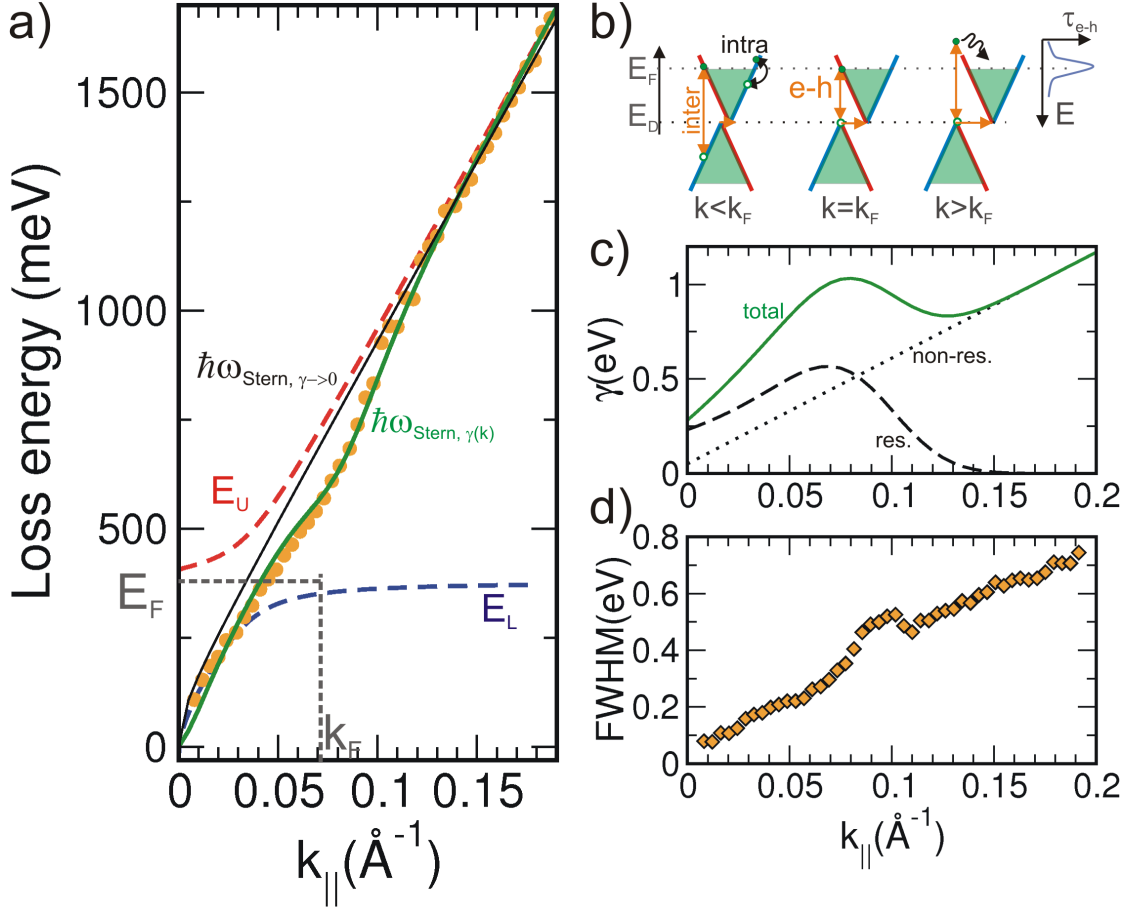}
\caption{\label{FIG3} (color online)
a) Fits of the plasmon dispersion
with the NFEG model including damping ($\hbar \omega_{Stern,\gamma (k)}$, thick(green) line) and
resonant plasmon electron-hole coupling ($E_U$ and $E_L$, coupling constant $2\Delta=100~meV$). For
$\Delta=0$ or damping constant $\gamma (k) \rightarrow 0$ the upper line  ($\hbar \omega_{Stern, \gamma  \rightarrow 0}$) is obtained.
b) Band diagram of
pristine epitaxial graphene at finite k--values demonstrating intra-band inter-band
e-h pair formation at finite $k_\parallel$ values. c) Effective damping $\gamma(k)$
and the decomposition into a resonant (res.) and
non-resonant (non-res.) part. d) The FWHM of the plasmon
loss is in qualitative agreement with $\gamma(k)$, shown in
c). For details, see text.}
\end{figure}

%Damping
In order to observe such resonance effects between excited states,
a short, but still sufficiently high lifetime
$\tau_{e-h}$ at resonance is necessary in order to discriminate both types as separate excitations.
As sketched in Fig.~\ref{FIG3}b) the finite value of $E_F$
above the Dirac point in pristine
graphene enables inter-band transitions at finite k-values,
which are shown as vertical transitions between shifted
conduction and valence band states \cite{Palmer87}.
Indeed, we can describe the whole dispersion curve in every detail
(see Fig.~\ref{FIG3}a) from the
loss function ($Im(-\nicefrac{1}{\epsilon})$), using simply
the Drude model of the dielectric function
$\epsilon=1-\nicefrac{\omega_p^2}{\omega(\omega +
i\gamma)}$, and the NFEG model of Stern assuming
a resonance by opening the $SPE_{inter}$ channel. This results
in  a momentum-dependent dissipation
$\gamma(k)$ for the plasmons.  The effective damping $\gamma(k)$ is
composed of a non--resonant term $\gamma_{non-res}$, which
takes into account intra-band transitions and structural imperfections, and a resonant
part $\gamma_{res}$. Fig.~\ref{FIG3}c) shows
the best fit result for $\gamma$. While $\gamma_{non-res}$
increases linearly with momentum \cite{Langer10}, $\gamma_{res}$ has a pronounced maximum around
$k_F$, as expected. The asymmetry in the damping curve
demonstrates further that intraband transitions at low
energies contribute to the dynamic loss channel as well.
This pronounced damping is clearly  seen in the experiment (Fig.~\ref{FIG3}d), although the increase of FWHM is smaller
than calculated. Finite k-resolution and the
simultaneous excitations of, e.g,  phonons, not considered here, may be possible reasons.
Nevertheless, the resonant broadening is even seen in much broader spectra, e.g.
in presence of F4-TCNQ molecules acting as defect scatterers.

%Polaritons
This purely phenomenological description of the dip can be rationalized
by a simple
two-oscillator model representing plasmon and e-h-pair formation by inter-band transitions
put on-top of the non-resonant decay mechanisms.  It yields a description fully consistent with the first approach.
In order to see this de-tuning effect   in the
dispersion, the  lifetimes, $\tau$, of both oscillators
need to be of  same order at $k_F$ . If further the
interaction strength $2\Delta \gg \hbar / \tau$, polaritons
are expected in this regime of strong coupling.  The upper and lower
polariton states are given by
$E_{U,L}(k)=\nicefrac{1}{2}(E_{pl}(k)+E_0) \pm
\nicefrac{1}{2} \sqrt{(2\Delta)^2+(E_{pl}(k)-E_0)^2}$,
where $E_{pl}$ and $E_0$ are the non--interacting plasmon
mode and exciton energy, respectively
\cite{Bellessa04,Agranovich03}. Here we set the sheet
plasmon energy $\hbar \omega_{Stern}$ for $E_{pl}$ and  the
Fermi energy $E_F$ for $E_0$, which is the lowest energy of
interband e-h pair formation without additional momentum.
$2\Delta$ measures the interaction between the quasi-particles.
The best fit  yields an interaction energy of
only 100~meV, in qualitative agreement with Ref.~\cite{Bostwick10}.
Consistent with the strongly enhanced damping determined from the fits described above,
the observation of separate Rabi split states is suppressed, and only the continuous
transition from the lower ($E_L$) to the upper branch ($E_U$) can be observed.
In fact, also this model fits the data quantitatively with values of damping close to
those determined above.
Formally, this model of plasmon electron-hole interaction
is closely related to interacting plasmons located in
spatially separated graphene layers \cite{Hwang09}, which
shows similar effects in the dispersion. However, since
smooth single graphene sheets were used here, we conclude
that the coupling to e-h oscillators is the dominant
process.

\begin{figure}[tb]
\includegraphics[width=0.75\columnwidth]{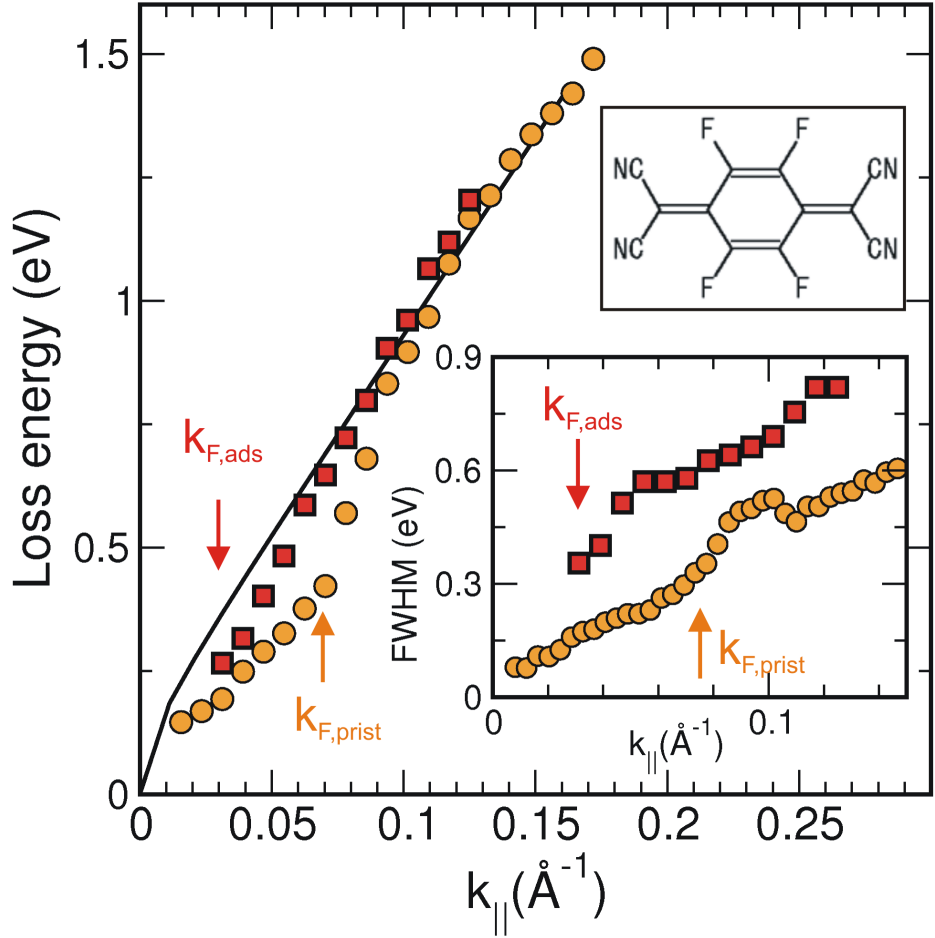}
\caption{\label{FIG4} (color online) Graphene plasmon dispersion before ($\bigcirc$)
and after F4-TCNQ doping ($\Box$). Inset: corresponding changes of
k-dependent FWHM. Upper inset: stereographic model of F4-TCNQ. Solid line:
Fit with the NFEG model without resonant damping.}
\end{figure}

% Adsorption of F4-TCNQ
Finally we show that the position of the dip in k-space is
only  influenced by the chemical potential rather than by
structural parameters. For this purpose, we adsorbed a
small amount of F4-TCNQ \cite{Chen07}, which reduces the carrier
density and hence shifts $E_F$
downwards. Fig.~\ref{FIG4} shows the dispersion after
adsorption of approximately 0.01~ML at 300~K. As expected,
the dip shifts to  lower $k_\parallel$-values ($\rm \approx
0.03~\AA^{-1}$).  This shift is also seen in the resonance
of the FWHM. Since the added molecules act as scattering
centers, the loss structure is further broadened so that
the exact determination of $k_F$ from the FWHM is more
difficult. Remarkably, the average slope of the dispersion
remains unchanged, although the electron density at $\rm E_F$
is reduced to $\rm 3 \times 10^{12} cm^{-2}$ due to adsorption
of F4-TCNQ, which may have two possible reasons: The
effective mass may be reduced by approximately the same
factor when reducing the filling of the bands above the
Dirac point, leaving N/m approximately constant.
Secondly, the coupling of the plasmon mode with the loss
channels, as evident from the large FWHM, leads to an
effective integration over the electron density around
$\rm E_F$ with a width proportional to the measured FWHM of the
plasmon mode. Both the dynamics of the single and the
collective excitation modes are extremely short and on the
same time scale, which results in mixing of collective and
single particle excitations.

%----Summary and conclusion-------------
In summary, the detailed analysis of plasmon excitations and their decay mechanisms,
using graphene as a 2d model system, yields ultra-short dynamics with life times of the order of $\rm \tau \approx 10^{-14}$s.  Both e-h pair excitations by intra- and inter-band
transitions contribute significantly to the decay of plasmons.  When the latter channel is opened
at finite $k_\parallel$, coupling between
plasmon and e-h pair excitations is resonantly enhanced.  As a consequence,
characteristic modifications of the plasmon dispersion and effective integration over
the density states at $E_F$ responsible for plasmon formation were found, thus widening the standard
picture for plasmon excitations.

We thank E.H. Hwang and S. Das Sarma for calculating the
plasmon dispersion of  graphene on the SiC dielectric (Fig. 2).

\bibliographystyle{h-physrev}

\end{document}